\documentclass[final,3p,twocolumn,sort&compress]{elsarticle}

\usepackage{amsmath}
\usepackage{amssymb}
\usepackage[figurename=Fig.,labelsep=period]{caption}

\journal{Physics Letters B}

\begin{document}

\begin{frontmatter}



\title{A two-dimensional soliton system of vortex and Q-ball}

\author[tpu]{A.Yu.~Loginov}
\ead{aloginov@tpu.ru}
\address[tpu]{Tomsk Polytechnic University, 634050 Tomsk, Russia}

\begin{abstract}
The $(2 + 1)$-dimensional  gauge  model  describing  two  complex scalar fields
that interact through a common Abelian gauge field is considered.
It is shown that  the  model has  a  soliton  solution  that describes a system
consisting of a vortex and a Q-ball.
This two-dimensional system  is electrically neutral, nevertheless it possesses
a nonzero electric field.
Moreover, the soliton system has a quantized magnetic flux and a nonzero angular
momentum.
Properties  of  this  vortex-Q-ball  system  are investigated by analytical and
numerical methods.
It is found that the system combines properties of topological and nontopological
solitons.

\end{abstract}


\end{frontmatter}

\section{Introduction}
\label{seq:I}

Topological solitons of $(2+1)$-dimensional field models play an important role
in field  theory,  physics  of  condensed  state, cosmology, and hydrodynamics.
First of all, it  is  necessary  to mention vortices of the effective theory of
superconductivity \cite{abr}  and vortices of the $(2 + 1)$-dimensional Abelian
Higgs model \cite{nielsen}.
Another  important  example  is  given  by  the  soliton  solution  of  the $(2
+1)$-dimensional nonlinear $O(3)$ $\sigma$ model \cite{belpol} that effectively
describes the behavior of a ferromagnet in the critical region.

Two-dimensional  soliton  solutions  of   Abelian   Maxwell  gauge  models  are
necessarily electrically neutral.
This is  because   the  $(2 + 1)$-dimensional Maxwell electrodynamics  does not
admit the existence of electrically charged  spatially localized solutions with
finite  energy  \cite{gladikowski}, in  contrast  to  the $(3 + 1)$-dimensional
case.
However,  the   electrical   neutrality   does   not  forbid  the  existence of
two-dimensional solitons possessing an electric field.

In this Letter we  consider  a  two-dimensional  soliton  system  consisting of
an Abelian vortex and a Q-ball.
The vortex and the Q-ball interact through a common Abelian gauge field.
This electrically  neutral  soliton  system  possesses a radial electric field,
carries a quantized magnetic flux, and  has  a nonzero angular momentum.
The soliton system combines the properties of vortex and Q-ball.
The interaction  between  the  vortex and the Q-ball by means of a common gauge
field leads to a significant change of their shapes.

\section{Lagrangian and field equations of the model}
\label{seq:II}

The $(2 + 1)$-dimensional  model  we  are  interested  in  is  described by the
Lagrangian density
\begin{eqnarray}
\mathcal{L} &=&-\frac{1}{4}F_{\mu \nu }F^{\mu \nu }+\left(D_{\mu }\phi
\right) ^{\ast }D^{\mu }\phi -V\left( \left\vert \phi \right\vert \right)
\nonumber \\
&&+\left( D_{\mu }\chi \right) ^{\ast }D^{\mu }\chi -U\left( \left\vert \chi
\right\vert \right),                                               \label{1}
\end{eqnarray}
where $\phi$ and $\chi$ are complex scalar fields that are minimally coupled to
the Abelian gauge field $A_{\mu}$ through covariant derivatives:
\begin{equation}
D_{\mu }\phi =\partial_{\mu }\phi -ieA_{\mu }\phi,\quad
D_{\mu }\chi =\partial_{\mu }\chi -iqA_{\mu }\chi.                    \label{2}
\end{equation}
The self-interaction potentials $V\left(\left\vert \phi \right\vert\right)$ and
$U\left(\left\vert \chi \right\vert\right)$ are
\begin{align}
V\left( \left\vert \phi \right\vert \right)  &=\frac{\lambda }{2}\left(
\phi ^{\ast }\phi -v^{2}\right) ^{2}\!, \nonumber \\
U\left( \left\vert \chi \right\vert \right)  &=m^{2}\chi ^{\ast }\chi
-g\left( \chi^{\ast }\chi \right)^{2}+
h \left( \chi^{\ast }\chi \right)^{3}\!,                              \label{3}
\end{align}
where $\lambda$, $g$, and $h$  are  the  positive  self-interaction  constants,
$m$ is the mass  of  the  scalar $\chi$-particle, and $v$ is the vacuum average
of the complex scalar field $\phi$.
We suppose that the potential $U\left( \left\vert \chi \right\vert \right)$ has
the global minimum at $\chi=0$ and a local one at some  $\left\vert \chi \right
\vert \ne 0$; hence  we  have  the  following condition for the parameters $m$,
$g$, and $h$:
\begin{equation}
\frac{g^{2}}{4m^{2}}<h<\frac{g^{2}}{3m^{2}}.                          \label{4}
\end{equation}
Note that if the coupling constant $q$ in Eq.~(2)  is  set  equal to zero, then
model (\ref{1}) has the soliton  solution  describing  an  Abelian vortex and a
two-dimensional Q-ball.
However, there is no electric field in this case,  so the vortex and the Q-ball
do not interact with each other.

The Lagrangian (\ref{1}) is invariant under the local gauge transformations:
\begin{eqnarray}
\phi \left( x\right)  &\rightarrow &\phi ^{\prime }\left( x\right) =\exp
\left( ie\Lambda \left( x\right) \right) \phi\left( x\right) , \nonumber \\
\chi \left( x\right)  &\rightarrow &\chi ^{\prime }\left( x\right) =\exp
\left( iq\Lambda \left( x\right) \right) \chi\left( x\right) , \nonumber \\
A_{\mu }\left( x\right)  &\rightarrow &A_{\mu }^{\prime }\left( x\right)
=A_{\mu }\left( x\right) +\partial _{\mu }\Lambda \left( x\right).    \label{5}
\end{eqnarray}
Moreover, the  Lagrangian (\ref{1}) is also invariant under the two independent
global gauge transformations:
\begin{eqnarray}
\phi \left( x\right)  &\rightarrow &\phi ^{\prime }\left( x\right) =\exp
\left( i\alpha \right) \phi \left( x\right) , \nonumber \\
\chi \left( x\right)  &\rightarrow &\chi ^{\prime }\left( x\right) =\exp
\left( i\beta \right) \chi \left( x\right).                           \label{6}
\end{eqnarray}
The corresponding Noether currents are
\begin{eqnarray}
j_{\phi }^{\mu } &=&-i\left[ \phi ^{\ast }D^{\mu }\phi -\left( D^{\mu }\phi
\right) ^{\ast }\phi \right], \nonumber \\
j_{\chi }^{\mu } &=&-i\left[ \chi ^{\ast }D^{\mu }\chi -\left( D^{\mu }\chi
\right) ^{\ast }\chi \right].  \label{7}
\end{eqnarray}

By varying the action $S = \int \mathcal{L}d^{3}x$ in $A_{\mu}$, $\phi^{\ast}$,
and $\chi^{\ast}$, we obtain the field equations of the model:
\begin{flalign}
& \partial_{\mu }F^{\mu \nu } = j^{\nu },                            \label{8a}
\\
& D_{\mu }D^{\mu }\phi +\lambda \left( \phi ^{\ast }\phi - v^{2}\right)
\phi = 0, \label{8b} \\
& D_{\mu }D^{\mu }\chi +\left( m^{2}-2g\left( \chi^{\ast }\chi \right)
+3h\left( \chi^{\ast }\chi \right)^{2}\right) \chi = 0,              \label{8c}
\end{flalign}
where the electromagnetic current $j^{\mu}$  is  expressed  in terms of Noether
currents:  
\begin{equation}
j^{\mu }=ej_{\phi }^{\mu }+qj_{\chi }^{\mu }.                         \label{9}
\end{equation}
Using the well-known  formula  $T_{\mu \nu }\!=\!2 \partial \mathcal{L}/\partial
g^{\mu \nu }\!-g^{\mu \nu }\mathcal{L}$, we obtain the symmetric energy-momentum
tensor of the model
\begin{align}
T_{\mu \nu } =&-F_{\mu \lambda }F_{\nu }^{\;\lambda }+\frac{1}{4}g_{\mu
\nu }F_{\lambda \rho }F^{\lambda \rho } \nonumber \\
&+\left( D_{\mu }\phi \right) ^{\ast }D_{\nu }\phi +\left( D_{\nu }\phi
\right) ^{\ast }D_{\mu }\phi  \nonumber \\
&-g_{\mu \nu }\left( \left( D_{\mu }\phi \right) ^{\ast }D^{\mu }\phi
-V\left( \left\vert \phi \right\vert \right) \right) \nonumber \\
&+\left( D_{\mu }\chi \right) ^{\ast }D_{\nu }\chi +\left( D_{\nu }\chi
\right) ^{\ast }D_{\mu }\chi  \nonumber \\
&-g_{\mu \nu }\left( \left( D_{\mu }\chi \right) ^{\ast }D^{\mu }\chi
-U\left( \left\vert \chi \right\vert \right) \right).                \label{10}
\end{align}
In particular, the energy density can be written as
\begin{align}
T_{00} = & \frac{1}{2}E_{i}E_{i}+\frac{1}{2}B^{2}              \label{11} \\
& +\left( D_{0}\phi \right) ^{\ast }D_{0}\phi + \left( D_{i}\phi \right)
^{\ast }D_{i}\phi +V\left( \left\vert \phi \right\vert \right) \nonumber  \\
& +\left( D_{0}\chi \right) ^{\ast }D_{0}\chi +\left( D_{i}\chi \right)
^{\ast }D_{i}\chi +U\left( \left\vert \chi \right\vert \right), \nonumber
\end{align}
where $E_{i} = F_{0i}$  are  the  components  of  electric  field  strength and
$B = -F_{12}$ is the magnetic field strength.

Let us fix the gauge as follows: $\partial _{0}\phi = 0$.
We want to find a soliton solution of model (\ref{1}) that minimizes the energy
functional $E = \int T_{00}d^{2}x$  at  the  fixed  value of the Noether charge
$Q_{\chi} = \int j_{\chi }^{0}d^{2}x$.
From the  method  of  Lagrange  multipliers  it  follows that the soliton is an
unconditional extremum of the functional
\begin{equation}
F=\int \mathcal{H}d^{2}x-\omega \int j_{\chi }^{0}d^{2}x =
E-\omega Q_{\chi},                                                   \label{14}
\end{equation}
where $\mathcal{H}$ is the Hamiltonian  density  and  $\omega$  is the Lagrange
multiplier.
Let us write the Noether charge $Q_{\chi}$ in terms of the canonically conjugated
variables:
\begin{equation}
Q_{\chi }=i\int \left( \chi \pi _{\chi }-\chi ^{\ast }\pi _{\chi ^{\ast
}}\right) d^{2}x,                                                    \label{15}
\end{equation}
where $\pi_{\chi }=\partial \mathcal{L}/\partial \left(\partial_{0}\chi \right)
=\left(D_{0}\chi \right) ^{\ast}$ and $\pi_{\chi ^{\ast }}= \partial \mathcal{L
}/\partial  \left( \partial _{0} \chi ^{\ast } \right)  = D_{0} \chi$  are  the
generalized  momenta   canonically  conjugated  to  $\chi$  and  $\chi^{\ast}$,
respectively.

The extremum condition for the functional $F$ is written as
\begin{equation}
\delta F = \delta H - \omega \delta Q_{\chi } = 0,                   \label{16}
\end{equation}
where a variation of the  Noether  charge $Q_{\chi}$ is written in terms of the
canonically conjugate variables:
\begin{equation}
\delta Q_{\chi } = i\int \left( \chi \delta \pi _{\chi }+\pi _{\chi }\delta
\chi -\text{c.c.}\right) d^{2}x.                                     \label{17}
\end{equation}
Using  the  Hamilton  field  equations  and  Eqs. (\ref{16}) and (\ref{17}), we
obtain:
\begin{equation}
\partial _{0}\chi = \frac{\delta H}{\delta \pi _{\chi }}=i\omega \chi
,\quad\partial _{0}\chi ^{\ast }=\frac{\delta H}{\delta \pi_{\chi^{\ast }}}
=-i\omega \chi^{\ast},                                               \label{18}
\end{equation}
while time derivatives of the  other model's fields are equal to zero.
From  Eq. (\ref{18}) we  get  the  time  dependence  of the scalar field $\chi$
\begin{equation}
\chi \left( x\right) = \chi \left( \mathbf{x}\right)
\exp \left( i \omega t \right),                            \label{19}
\end{equation}
whereas  the other  fields  of  the  model  do  not depend on time in the gauge
$\partial_{0} \phi = 0$.
From Eq. (\ref{16}) it  follows  that  the  important  relation  holds  for the
soliton solution:
\begin{equation}
\frac{dE}{dQ_{\chi }} = \omega,                                      \label{20}
\end{equation}
where $\omega$ is some function of $Q_{\chi}$.

\section{The ansatz and some properties of the solution}
\label{seq:III}

To find the soliton  solution  of  field  equations (\ref{8a}), (\ref{8b}), and
(\ref{8c}), we use the following ansatz for the model's fields:
\begin{align}
A^{\mu }\left( x\right) &= \left( \frac{A_{0}\left( r\right) }{e r},\frac{1}
{e r} \epsilon_{ij} n_{j} A\left( r\right) \right) , \nonumber \\
\phi \left( x\right)  & = v\exp \left( - i N \theta \right) F\left( r\right),
\nonumber \\
\chi \left( x\right) & = \sigma \left( r\right) \exp \left( i\omega t\right),
                                                                     \label{21}
\end{align}
where $\epsilon_{ij}$  and  $n_{j}$  are  the components of the two-dimensional
antisymmetric tensor $\left(\epsilon_{12}=1 \right)$ and the radial unit vector
$\mathbf{n} =\left(\cos\left(\theta\right)\!,\,\sin\left(\theta\right)\right)$,
respectively.
Note that the fields $A^{i}$ and $\phi$ are described by the vortex ansatz that
was used  in  \cite{hong}, while  the  scalar  field $\chi$ is described by the
Q-ball ansatz \cite{coleman}.
Note also that ansatz (\ref{21}) completely fixes the model's gauge.

Substituting ansatz (\ref{21}) into field equations (\ref{8a}), (\ref{8b}), and
(\ref{8c}),  we  obtain the  system  of ordinary differential equations for the
ansatz functions $A_{0}(r)$, $A(r)$, $F(r)$, and $\sigma(r)$:
\begin{eqnarray}
&& A_{0}^{\prime \prime }(r)-\frac{A_{0}^{\prime }(r)}{r}+\frac{A_{0}(r)}
{r^{2}} \nonumber \\
&& -\left( 2 e^{2}v^{2}F\left( r\right) ^{2}+2q^{2}\sigma \left( r\right)
^{2}\right)  \nonumber \\
&& \times A_{0}(r) + 2 e q \omega r \sigma \left( r \right)^{2} = 0,
                                                                     \label{22}
\end{eqnarray}
\begin{eqnarray}
&& A^{\prime \prime }(r)-\frac{A^{\prime }(r)}{r}-2e^{2}v^{2}
\left(N + A(r)\right) \nonumber \\
&& \times F\left( r\right)^{2}-2q^{2}\sigma \left( r\right)^{2}
A\left(r\right) = 0,                                                 \label{23}
\end{eqnarray}
\begin{eqnarray}
&& F^{\prime \prime }(r)+\frac{F^{\prime }(r)}{r} -
\frac{F(r)}{r^{2}} \nonumber \\
&& \times \left((N + A(r))^{2}-A_{0}(r)^{2}\right) \nonumber \\
&& +\lambda v^{2}\left( 1-F(r)^{2}\right) F(r) = 0,                  \label{24}
\end{eqnarray}
\begin{eqnarray}
&& \sigma ^{\prime \prime }(r)+\frac{\sigma ^{\prime }(r)}{r}+
\sigma \left( r\right) \nonumber  \\
&& \times \left( \left(\omega - \frac{q}{e}\frac{A_{0}
\left( r\right)}{r}\right)^{2}-\frac{q^{2}}{e^{2}}
\frac{A\left( r\right) ^{2}}{r^{2}}\right)  \label{25} \\
&& -\left( m^{2}-2g\sigma \left( r\right) ^{2}+3h\sigma
\left( r\right)^{4}\right) \sigma \left( r\right)  = 0.  \nonumber
\end{eqnarray}

Substituting ansatz (\ref{21}) into  Eq. (\ref{11}),  we  obtain the expression
for the energy density in terms of the ansatz functions:
\begin{flalign}
\mathcal{E} =& \,\frac{A^{\prime }{}^{2}}{2e^{2}r^{2}}+\frac{1}{2}
\left( \left( \hspace{-0.03in}\frac{A_{0}}{er}\hspace{-0.03in}\right)
^{\prime }\right) ^{2}+v^{2}F^{\prime }{}^{2}  \nonumber \\
& +\frac{\left( (N + A)^{2}+A_{0}{}^{2}\right) }{r^{2}}
v^{2}F^{2}                                     \nonumber \\
&+\frac{\lambda }{2}v^{4}\left( F^{2}-1\right) ^{2}+\sigma ^{\prime }{}^{2}
                                               \nonumber \\
&+\left( \omega -q\frac{A_{0}}{er}\right) ^{2}\sigma ^{2}+\frac{q^{2}}{e^{2}
}\frac{A^{2}}{r^{2}}\sigma ^{2}                \nonumber \\
&+m^{2}\sigma ^{2}-g\sigma ^{4}+h\sigma^{6}.                         \label{26}
\end{flalign}
It follows from  the  regularity  condition  of the soliton solution at $r = 0$
and from the finiteness of the soliton's energy $E=2\pi\int\nolimits_{0}^{\infty}
\mathcal{E}\left(r\right)r dr$  that the ansatz functions satisfy the following
boundary conditions:
\begin{align}
A_{0}(0) &= 0, \quad A_{0}(r) \underset{r\rightarrow \infty }{\longrightarrow}0,
\nonumber \\
A(0) &= 0, \quad \, A(r) \underset{r\rightarrow \infty }{\longrightarrow } -N,
\nonumber \\
F(0) &= 0, \quad \, F(r) \underset{r\rightarrow \infty }{\longrightarrow }1,
\nonumber \\
\sigma^{\prime }(0) &= 0, \quad \, \sigma (r) \underset{r\rightarrow \infty}
{\longrightarrow } 0.                                                \label{27}
\end{align}
The boundary conditions for $A(r)$ lead to  the  magnetic flux quantization for
the vortex-Q-ball system
\begin{equation}
\Phi = 2 \pi \int_{0}^{\infty }B\left(r\right) rdr=\frac{2\pi }{e}N, \label{28}
\end{equation}
where $B(r)=-A^{\prime }(r)/(er)$ is the magnetic field strength.

Substituting   the   power   expansions  for  $A_{0}(r)$,  $A(r)$,  $F(r)$, and
$\sigma(r)$  into  Eqs. (\ref{22})--(\ref{25}) and taking into account boundary
conditions  (\ref{27}),  we  obtain  the  asymptotic form  of  the  solution as
$r \rightarrow 0$:
\begin{align}
A_{0}\left( r\right)& =a_{1}r+\frac{a_{3}}{3!}r^{3}+O\left( r^{5}\right),
\nonumber \\
A\left( r\right)& =\frac{b_{2}}{2!}r^{2}+\frac{b_{4}}{4!}r^{4}+
O\left(r^{6}\right),
\nonumber \\
F\left( r\right)& =\frac{c_{\left|N\right|}}{\left|N\right|!}r^{\left|N\right|}
\!+\!\frac{c_{\left|N\right|+2}}{\left( \left|N\right|\!+\!2\right)!}
r^{\left|N\right|+2}\!+\!O\!\left(\!r^{\left|N\right|+4}\!\right)\!,
\nonumber \\
\sigma \left( r\right)& =d_{0}+\frac{d_{2}}{2!}r^{2}+O\left( r^{4}\right).
\label{28a}
\end{align}
In  Eq.~(\ref{28a}),   the   next-to-leading   coefficients  $a_{3}$,  $b_{4}$,
$c_{\left| N \right| + 2}$,  and  $d_{2}$ are expressed in terms of the leading
coefficients and the model's parameters:
\begin{align}
a_{3} = & \, 3 q d_{0}^{2}\left( a_{1}q-e\omega \right), \nonumber \\
b_{4} = & \, 3 \left( q^{2}b_{2}d_{0}^{2}+2 N e^{2}v^{2}c_{\left|N\right|}^{2}
\delta_{1,\left|N\right|}\right), \nonumber \\
c_{\left|N\right|+2} = & -\frac{c_{\left|N\right|}}{4}
\left(\left|N\right|+2\right)\left( a_{1}^{2}+\left|N\right|\left|b_{2}\right|
+\lambda v^{2}\right), \nonumber \\
d_{2} = & \frac{d_{0}}{2}\left[ d_{0}^{2}\left( 3d_{0}^{2}h-2g\right) \right.
\nonumber \\
&\left. +e^{-2}\left( qa_{1}+e\left( m-\omega \right) \right) \right.
\nonumber \\
&\left. \times \left( -qa_{1}+e\left( m+\omega \right) \right) \right],
\label{28b}
\end{align}
where $\delta_{1,\left|N\right|}$ is the Kronecker symbol.
Linearization  of  Eqs.  (\ref{22})--(\ref{25})  at  large  $r$  together  with
corresponding boundary conditions (\ref{27}) lead  us to the asymptotic form of
the solution as $r \rightarrow \infty$:
\begin{align}
A_{0}\left( r\right)  &\sim a_{\infty }\sqrt{m_{A}r}\exp \left(
-m_{A}r\right), \nonumber \\
A\left( r\right)  &\sim - N + b_{\infty }\sqrt{m_{A}r}\exp \left(
-m_{A}r\right), \nonumber \\
F\left( r\right)  &\sim 1 + c_{\infty}\frac{\exp \left(-m_{\phi }r\right)}
{\sqrt{m_{\phi }r}}, \nonumber \\
\sigma \left( r\right)&  \sim  d_{\infty }\frac{\exp
\left(-\sqrt{m^{2}-\omega^{2}} r \right) }
{\sqrt{\sqrt{m^{2}-\omega^{2}} r}},                                 \label{28c}
\end{align}
where $m_{A} = \sqrt{2} e v$ and $m_{\phi} = \sqrt{2 \lambda} v$ are the masses
of the gauge boson and the scalar $\phi$-particle, respectively.

Eq.~(\ref{22}) can be rewritten in the compact form 
\begin{equation}
-\left( r\left( \frac{A_{0}\left( r\right) }{er}\right) ^{\prime }\right)
^{\prime }=rj_{0}\left( r\right),                                   \label{28d}
\end{equation}
where the  zero  component  $j_{0}$  of  electromagnetic  current  (\ref{9}) is
written in terms of the ansatz  functions
\begin{equation}
j_{0}=2q\omega \sigma ^{2}-\frac{2A_{0}}{er}\left( q^{2}\sigma
^{2}+e^{2}v^{2}F^{2}\right).                                        \label{28e}
\end{equation}
Let us integrate the both sides of Eq.~(\ref{28d}) with respect to $r$ from $0$
to $\infty$.
Taking  into  account  boundary  conditions  (\ref{27})  and  asymptotic  forms
(\ref{28a}) and (\ref{28c}), it  can  be  easily  shown  that  the  integral of
the left-hand side of Eq.~(\ref{28d}) vanishes.
At the same  time, the  integral  of  the right-hand side of Eq.~(\ref{28d}) is
equal to $Q/(2\pi)$, where $Q$ is the soliton's electric charge.
Hence the soliton's electric charge is equal to zero.
This fact and Eq.~(\ref{9}) lead us to the relation between the Noether charges
$Q_{\phi } =\int j_{\phi }^{0}d^{2}x$ and $Q_{\chi }=\int j_{\chi }^{0} d^{2}x$
\begin{equation}
Q = eQ_{\phi }+qQ_{\chi } = 0.                                       \label{13}
\end{equation}

In the case of symmetric energy-momentum tensor (\ref{10}), the angular momentum
tensor is written as
\begin{equation}
J^{\lambda \mu \nu }=x^{\mu }T^{\lambda \nu }-x^{\nu }T^{\lambda \mu }.
                                                                     \label{29}
\end{equation}
From Eqs.~(\ref{10}), (\ref{21}), and (\ref{29}), we  obtain  the expression of
the angular momentum's density in terms of the ansatz functions:
\begin{align}
\mathcal{J} &=\frac{1}{2}\epsilon _{ij}J^{0ij}=- r B E_{r}+2\frac{q}{e}A\left(
\omega - q\frac{A_{0}}{er}\right) \sigma ^{2} \nonumber \\
&-2\frac{A_{0}\left( N+A\right) }{r}v^{2}F^{2},                      \label{30}
\end{align}
where $ E_{r}(r)=-\left( A_{0}\left(r\right)/\left( er\right) \right)^{\prime}$
is the radial component of the electric field strength.
Integrating the term $-rBE_{r}=-e^{-2}A^{\prime }\left(A_{0}/r\right)^{\prime}$
by parts, taking  into  account  boundary  conditions (\ref{27}), and using Eq.
(\ref{22}) to  eliminate  $A_{0}^{ \prime \prime}$,  we  obtain  the  following
expression for the angular momentum $J=2\pi \int_{0}^{\infty }\mathcal{J}\left(
r\right)rdr$:
\begin{equation}
J = -4 \pi N v^{2} \int_{0}^{\infty }A_{0}\left( r\right)F^{2}(r)dr. \label{31}
\end{equation}
From Eqs.~(\ref{7}) and (\ref{21}) it follows that the Noether charge $Q_{\phi}$
can be written in terms of the ansatz functions as
\begin{equation}
Q_{\phi }=-4\pi v^{2}\int_{0}^{\infty}A_{0}\left(r\right)F^{2}(r)dr. \label{32}
\end{equation}
Comparing Eqs.~(\ref{31}) and (\ref{32}), and taking into account Eq.~(\ref{13}),
we obtain the important  relation  between  the  angular  momentum $J$  and the
Noether charges $Q_{\phi}$ and $Q_{\chi}$ of the vortex-Q-ball system:
\begin{equation}
J = N Q_{\phi} = -\frac{q}{e} N Q_{\chi}.                            \label{33}
\end{equation}

Any solution of field equations (\ref{8a}) -- (\ref{8c}) is an  extremum of the
action $S=\int \mathcal{L}d^{2}xdt$.
Hovever, for the field configurations of ansatz (\ref{21}), Lagrangian  density
(\ref{1}) does not depend on time.
Consequently, any  solution  of the system of differential equations (\ref{22})
-- (\ref{25}) is  an  extremum  of the Lagrangian $L = \int \mathcal{L}d^{2}x$.
Let $A_{0}(r)$,  $A(r)$,  $F(r)$,  and  $\sigma(r)$  be  a  solution  of system
(\ref{22}) -- (\ref{25}) satisfying boundary conditions (\ref{27}).
Let us  perform  the  scale  transformations  of  the  solution's  argument: $r
\rightarrow \lambda r$.
After  that, the  Lagrangian $L$  becomes  a  function  of  the scale parameter
$\lambda$.
Equating to zero the derivative $ dL/d\lambda $ at $\lambda = 1$, we obtain the
virial relation for the vortex-Q-ball system:
\begin{equation}
E^{\left( E\right) }-E^{\left( B\right) }+E^{\left( P\right) }-\frac{\omega
}{2}Q_{\chi } = 0,                                                   \label{34}
\end{equation}
where
\begin{equation}
E^{\left( E\right) }=\frac{1}{2}\int E_{i}E_{i} d^{2}x =
\pi \int_{0}^{\infty}\left( \left( \hspace{-0.03in}
\frac{A_{0}}{er}\hspace{-0.03in}\right)^{\prime }\right)^{2}r dr     \label{35}
\end{equation}
is the energy of the electric field,
\begin{equation}
E^{\left( B\right) }=\frac{1}{2}\int B^{2}d^{2}x =
\pi \int_{0}^{\infty} \frac{A^{\prime}{}^{2}}{e^{2}r} dr             \label{36}
\end{equation}
is the energy of the magnetic field,  and
\begin{equation}
E^{\left( P\right) }=2\pi\int_{0}^{\infty}
\left[V\left( \left\vert \phi \right\vert \right)
+U\left( \left\vert \chi \right\vert \right) \right] r dr            \label{37}
\end{equation}
is the potential part of the soliton's energy.

\section{Numerical results}
\label{seq:IV}

\begin{figure}[t]
\includegraphics[width=7.8cm]{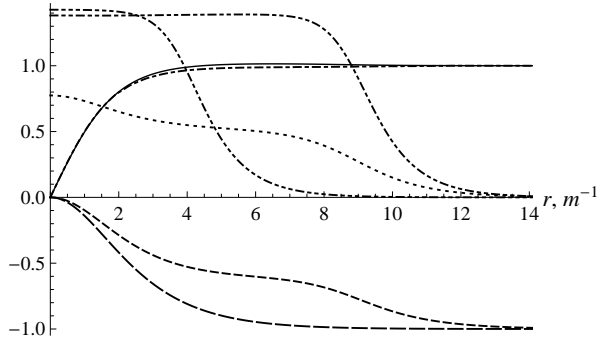}
\caption{\label{fig1}  The  numerical  solution  for  the  dimensionless ansatz
functions  $m^{-1/2}A_{0}(r)/(er)$  (dotted),  $A(r)$   (short-dashed),  $F(r)$
(solid), and $m^{-1/2} \sigma(r)$ (dash-dot-dotted) of the vortex-Q-ball system
and  for  the  dimensionless  ansatz  functions  $A(r)$  (long-dashed),  $F(r)$
(dash-dotted),    and    $m^{-1/2}\sigma(r)$   (dash-dot-dot-dotted)   of   the
noninteracting vortex and Q-ball.  The  model's  parameters  are the following:
$e = q = 0.3\,m^{1/2}$, $\lambda = 0.173\,m$, $v = 1.7\,m^{1/2}$, $g = 1.0\,m$,
$h = 0.26$,  and $N=1$. The phase frequency $\omega = 0.4\,m$.}
\end{figure}
Now let us present some numerical results.
We use the natural units $c = 1$,  $\hbar = 1$, and  the  mass  $m$  of  scalar
$\chi$-particle  is used as the energy unit.
Then the model depends  on  the  six parameters: $e$, $q$, $\lambda$, $v$, $g$,
and $h$.
Let us choose the following values of these parameters: $e = q = 0.3\,m^{1/2}$,
$\lambda = 0.173\,m$, $v = 1.7\,m^{1/2}$,  $g = 1.0\,m$, and  $h = 0.26$, where
the  parameters'  dimensions  correspond  to  the $(2+1)$-dimensional case.
Such a choice corresponds to the masses $m_{\phi }=\sqrt{2 \lambda }v = 1.0\,m$
and $m_{A} = \sqrt{2} e v = 0.72\,m$  of  the  scalar  $\phi$-particle  and the
gauge boson, respectively.
Note  that  the  mass  ratio  $m_{A}/m_{\phi}$  is  close  to  the  mass  ratio
$m_{Z}/m_{H}$ of the Standard model.
To check the correctness  of  numerical  solution, Eqs. (\ref{20}), (\ref{13}),
(\ref{33}), and (\ref{34}) were used.

Figure 1 presents the  numerical  solution for the dimensionless zero component
$m^{-1/2}A_{0}(r)/(er)$ of the gauge potential and for the dimensionless ansatz
functions $A(r)$, $F(r)$, and $m^{-1/2}\sigma(r)$.
The vortex part of the solution is in the topological sector  with $N = 1$, the
phase frequency $\omega$ is equal to $0.4\,m$.
Figure 1 also presents the numerical solution for the case $q = 0$, whereas the
other parameters remain the same.
The case $q = 0$ corresponds  to  superimposed  but  noninteracting  vortex and
Q-ball.
From Fig.~1 it follows that the interaction  between  the vortex and the Q-ball
has a significant  effect  on  the  shapes  of  the ansatz functions $A(r)$ and
$\sigma(r)$, while the shape of $F(r)$ does not change significantly.

\begin{figure}[t]
\includegraphics[width=7.8cm]{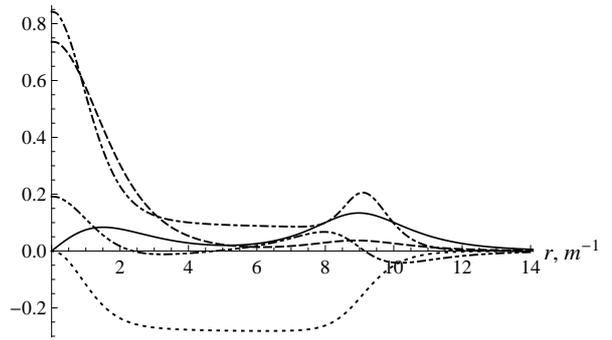}
\caption{\label{fig2} The dimensionless versions of the electric field strength
$\tilde{E}_{r}(r)  =  m^{-3/2}E_{r}(r)$ (solid), the  magnetic  field  strength
$\tilde{B}(r)  =  m^{-3/2}B(r)$  (dashed), the  scaled  energy  density  $0.3\,
\tilde{\mathcal{E}}(r) = 0.3\,m^{-3}\mathcal{E}(r)$ (dash-dotted), the electric
charge density $\tilde{j}_{0}(r) = m^{-5/2}j_{0}(r)$ (dash-dot-dotted), and the
scaled angular  momentum's  density  $0.3\,\tilde{\mathcal{J}}(r) = 0.3\,m^{-2}
\mathcal{J}(r)$ (dotted), corresponding to the solution in Fig.~1.}
\end{figure}
Figure 2 shows  the  dimensionless  versions  of  the  electric  field strength
$\tilde{E}_{r}(r)=m^{-3/2}E_{r}(r)$, the magnetic field strength $\tilde{B}(r)=
m^{-3/2}B(r)$, the scaled  energy  density $0.3\,\tilde{\mathcal{E}}(r) = 0.3\,
m^{-3}\mathcal{E}(r)$, the electric charge  density  $\tilde{j}_{0}(r)=m^{-5/2}
j_{0}(r)$, and  the scaled angular momentum's density $0.3\,\tilde{\mathcal{J}}
(r) = 0.3\,m^{-2} \mathcal{J}(r)$, corresponding to the solution in Fig.~1.
From Fig.~2 it  follows  that  the  vortex-Q-ball system can roughly be divided
into the three parts: the central transition region,  the inner region, and the
external transition region.
In the inner region, the  energy density and the angular momentum's density are
approximately constant, while  the  electric  and  magnetic field strengths are
close to zero.

\begin{figure}[t]
\includegraphics[width=7.8cm]{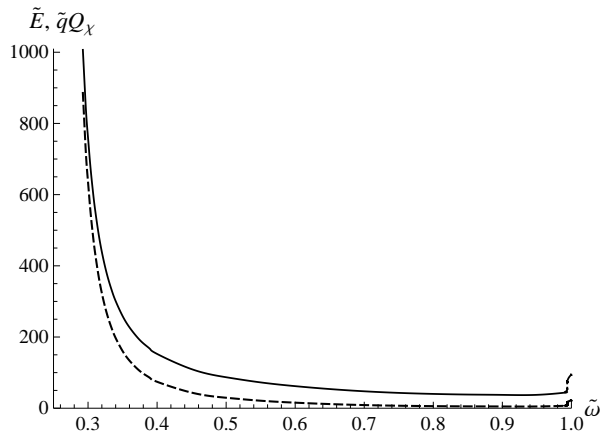}
\caption{\label{fig3} The  dependences  of  the  dimensionless  soliton  energy
$\tilde{E} = m^{-1}E$ (solid) and  the  dimensionless  electric charge $\tilde{
q}Q_{\chi} = m^{-1/2} q Q_{\chi}$ (dashed)  of  the  scalar field $\chi$ on the
dimensionless  phase  frequency $\tilde{\omega} = m^{-1}\omega$.   The  model's
parameters are the same as in Fig.~1.}
\end{figure}
Figure  3  presents  the   dependences  of  the  dimensionless  soliton  energy
$\tilde{E} = m^{-1}E$ and the dimensionless electric  charge $\tilde{q}Q_{\chi}
= m^{-1/2} q Q_{\chi}$ of the scalar field $\chi$  on  the  dimensionless phase
frequency  $\tilde{\omega} = m^{-1}\omega$.
The  dependences  are  presented  in  the  range  from  the  minimum  value  of
$\tilde{\omega}$ to  its  maximum  value  of  $1$  that  we managed to reach by
numerical methods.
From Fig.~3 it follows that the energy $E$ and the Noether charge $Q_{\chi}$ of
the vortex-Q-ball system tend to infinity as $\tilde{\omega}\rightarrow \tilde{
\omega}_{\min}$ (thin-wall regime).
In the  thin-wall  regime, the  spatial  size  of  the  soliton's  inner region
increases indefinitely, so the  main contribution to the soliton's energy comes
from this region.

In Fig.~4 we can  see the dependences that are the same as those in Fig.~3, but
are shown in a neighborhood of the maximum value $\tilde{\omega} = 1$.
From  Fig.~4  it  follows  that  the  curves  $\tilde{E}( \tilde{\omega} )$ and
$Q_{\chi}(\tilde{\omega})$ consist of two branches.
The left branches are finished at $\tilde{\omega}_{-} \approx 0.99404$, whereas
the right ones are started at $\tilde{\omega}_{+} \approx 0.99389$.
Note that $\tilde{\omega}_{-} > \tilde{\omega}_{+}$  so  that  the branches are
overlapped.
It was found numerically that $Q_{\chi}(\tilde{\omega})$ and $\tilde{E}(\tilde{
\omega})$ have the following  behaviour  as $\tilde{\omega} \rightarrow \tilde{
\omega}_{\pm }$:
\begin{align}
&Q_{\chi }\underset{\tilde{\omega} \rightarrow \tilde{\omega}_{\pm }}
{\longrightarrow}Q_{\chi_{\pm}}\pm 3A_{\pm}\left(\mp\tilde{\omega}_{\pm}
\pm \tilde{\omega}\right)^{\frac{1}{3}},                             \label{38}
\\
&\tilde{E}\underset{\tilde{\omega} \rightarrow \tilde{\omega}_{\pm }}
{\longrightarrow }\tilde{E}_{\pm }\pm\frac{3}{4}A_{\pm }
\left(3\tilde{\omega}_{\pm }+\tilde{\omega} \right) \left( \mp \tilde{\omega}
_{\pm }\pm \tilde{\omega} \right) ^{\frac{1}{3}},  \nonumber
\end{align}
where $A_{\pm}$ are positive constants.
Note that  the  behaviour  of $Q_{\chi}(\tilde{\omega})$ and $\tilde{E}(\tilde{
\omega})$ in neighborhoods of $\tilde{\omega}_{+}$  and $\tilde{\omega}_{-}$ is
in agreement with Eq.~(\ref{20}).
From Eq.~(\ref{38}) it follows that the left and right branches have the branch
points at $\tilde{\omega}_{-}$ and $\tilde{\omega}_{+}$, respectively.
Such  behaviour of $Q_{\chi}(\tilde{\omega})$  and  $\tilde{E}(\tilde{\omega})$
in a neighborhood of the  maximum  value $\tilde{\omega} = 1$ is very different
from that of the two-dimensional Q-ball \cite{lee}.
\begin{figure}[t]
\includegraphics[width=7.8cm]{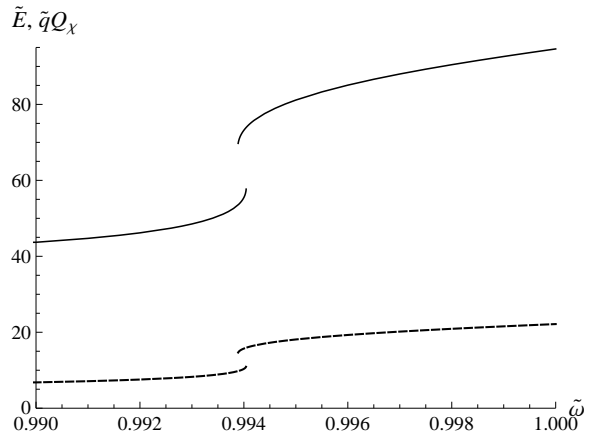}
\caption{\label{fig4}   The same dependences as those in Fig.~3, but shown in a
neighborhood of $\tilde{\omega} = 1$.}
\end{figure}

Figure 5  shows  the  dependence  of  the  vortex-Q-ball system's dimensionless
energy $\tilde{E}$ on the Noether charge $Q_{\chi}$.
It also shows the similar dependence  for  the  two-dimensional Q-ball with the
same parameters $m$, $g$, and $h$ as the vortex-Q-ball system, and the straight
line $\tilde{E} = Q_{\chi}$.
We can  see  that  the  two-dimensional Q-ball's curve $\tilde{E}(Q_{\chi})$ is
tangent to the straight line $\tilde{E} = Q_{\chi}$ as it should be \cite{lee}.
In contrast to this,  the  vortex-Q-ball  system's  curve $\tilde{E}(Q_{\chi})$
has the cusp.
Moreover this curve has  the  gap that corresponds to the jump from the left to
the right branches in Fig.~4.
From Fig.~5 it  follows  that  the  Q-ball  component  of the the vortex-Q-ball
system is stable to the  decay  in  the  massive scalar $\chi$-particles in the
thin-wall regime.
\begin{figure}[t]
\includegraphics[width=7.8cm]{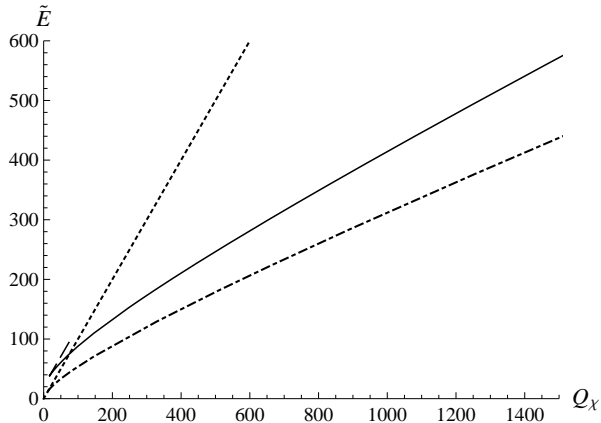}
\caption{\label{fig5} The dependence of the vortex-Q-ball system's dimensionless
energy $\tilde{E}$ on the Noether charge $Q_{\chi}$  (solid)  and  that  of the
two-dimensional Q-ball (dash-dotted) with the same parameters $m$, $g$, and $h$
as the vortex-Q-ball system.  The  dotted  line is the straight line $\tilde{E}
= Q_{\chi}$.}
\end{figure}

\section{Conclusions}
\label{seq:V}

In the present paper, the soliton  system  consisting  of a vortex and a Q-ball
interacting through a common Abelian gauge field has been researched.
This two-dimensional system  is  electrically  neutral  since  only the Maxwell
gauge term is presented in the Lagrangian (\ref{1}).
Nevertheless, the  vortex-Q-ball  system  possesses  a  nonzero radial electric
field.
Moreover, this system also has a quantized magnetic flux.
As a result, the soliton system possesses a nonzero angular momentum that turns
out to  be  proportional  to  the  Noether  charge  of  the  scalar  $\phi$- or
$\chi$-field.
The  vortex-Q-ball  system  combines   properties  of  nontopological  solitons
(Eq.~(\ref{20}))  and  those  of  topological  solitons  (topological  boundary
condition (\ref{27}) for $A(r)$ and, as consequence, magnetic flux quantization
(\ref{28})).
Finally,  the  interaction  between  the  vortex  and  the  Q-ball leads to the
significant  change  of  the  vortex-Q-ball  system's dependence  $E(Q_{\chi})$
in comparison with that of the two-dimensional Q-ball.

It should be noted that in $(2 + 1)$  dimensions, in  addition  to the ordinary
Lorentz-invariant Maxwell term,  there exists  a Lorentz-invariant Chern-Simons
term, which can be included in Lagrangians of gauge models \cite{JT, schonfeld,
 DJT}.
In the presence of this term,  the  model's gauge  field  becomes topologically
massive, thus  making  possible  the existence of two-dimensional solitons that
have a nonzero  quantized  electric charge \cite{hong, paul, paul2, ghosh, jw1,
 jw2}.
Due to the  presence  of  electric  and  magnetic  fields,  these solitons also
possess   nonzero  angular  momentums  that  satisfy  the  relations similar to
Eq.~(\ref{33}).

\section*{Acknowledgments}

The research  is  carried  out  at  Tomsk  Polytechnic  University  within  the
framework of  Tomsk  Polytechnic University Competitiveness Enhancement Program
grant.





\bibliographystyle{elsarticle-num}

\bibliography{article}






\end{document}